\documentclass[journal=acsnano, manuscript=article]{achemso}

\usepackage{chemformula} 
\usepackage[T1]{fontenc} 
\usepackage{siunitx}
\usepackage{mathtools}
\usepackage{xcolor}
\usepackage[markup=underlined]{changes}
\usepackage{xr-hyper}
\setkeys{acs}{maxauthors=20}
\setkeys{acs}{etalmode=truncate}

\makeatletter
\newcommand*{\addFileDependency}[1]{
  \typeout{(#1)}
  \@addtofilelist{#1}
  \IfFileExists{#1}{}{\typeout{No file #1.}}
}
\makeatother

\author{Woo Je Chang}
\affiliation[UTChemE]{McKetta Department of Chemical Engineering, University of Texas at Austin, Austin, Texas 78712, United States}
\author{Allison M. Green}
\affiliation[UTChemE]{McKetta Department of Chemical Engineering, University of Texas at Austin, Austin, Texas 78712, United States}
\author{Zarko Sakotic}
\affiliation[UTEECS]{Chandra Family Department of Electrical and Computer Engineering, University of Texas at Austin, Austin, Texas 78758, USA}
\author{Daniel Wasserman}
\affiliation[UTEECS]{Chandra Family Department of Electrical and Computer Engineering, University of Texas at Austin, Austin, Texas 78758, USA}
\email{dw@utexas.edu}
\author{Thomas M. Truskett}
\affiliation[UTChemE]{McKetta Department of Chemical Engineering, University of Texas at Austin, Austin, Texas 78712, United States}
\alsoaffiliation[UTPhys]{Department of Physics, University of Texas at Austin, Austin, Texas 78712, United States}
\email{truskett@che.utexas.edu}
\author{Delia J. Milliron}
\affiliation[UTChemE]{McKetta Department of Chemical Engineering, University of Texas at Austin, Austin, Texas 78712, United States}
\alsoaffiliation[UTChem]{Department of Chemistry, University of Texas at Austin, Austin, Texas 78712, United States}
\email{milliron@che.utexas.edu}

\title{Plasmonic metal oxide nanocrystals as building blocks for
infrared metasurfaces}

\begin{document}

\begin{tocentry}
\includegraphics[width=\textwidth]{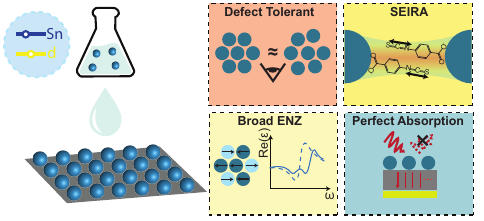}
\end{tocentry}

 \renewcommand{\abstractname}{Conspectus}
\begin{abstract}

Metamaterials operating at infrared (IR) frequencies have garnered significant attention due to the opportunities for resonant interactions with vibrational modes of molecules and materials, and manipulation of thermal emission. These metamaterials usually consist of periodic arrangements of sub-wavelength scale metallic or dielectric elements, patterned either top-down by nanolithographic methods or bottom-up by nanocrystal (NC) assembly. However, conventional metals are inherently constrained by their fixed electron concentrations, which limits the degrees of freedom in the design of the meta-atom unit cells to achieve the desired optical response. In this context, doped metal oxide NCs, with the prototypical case being tin-doped indium oxide (ITO) NCs, are an exceptional candidate for self-assembled IR metamaterials, owing to their relatively low and synthetically tunable electron concentrations that govern the frequencies of their IR plasmon resonances. Focusing on ITO NCs as building blocks, this Account describes recent progress in synthetic tuning of NC optical properties, NC superlattice monolayer preparation methods for fabricating IR resonant metamaterials, and the emerging understanding of the optical response, facilitated by recently developed simulation methods.

Based on experimental and simulation methods we helped develop, we are advancing mechanistic understanding of how self-assembled NC metamaterials can produce distinctive near- and far-field optical properties not readily achievable in lithographically patterned structures. First,  the impacts of the inevitable defects and disorder associated with self-assembly can be rationalized and, in some cases, recognized as advantageous. Second, self-assembly enables intimate nanoscale intermixing of different NC and molecular components. By incorporating probe molecules within the gaps between NCs where the electric field enhancement is strongest, we show enhanced detection of molecular vibrations that can be optimized by tuning the size and resonance frequency of the NCs. We show how metasurfaces incorporating mixtures of NCs with different doping concentrations can achieve epsilon-near-zero dielectric response over a broad frequency range. Finally, considering the NC metasurface itself as a building block, we show how photonic structures incorporating these assemblies can harness and amplify their distinctive properties. Through modeling the NC monolayer as a slab with an effective permittivity response, we designed a frequency-tunable IR perfect absorber by layering the NCs on a simple open cavity structure. Since the perfect absorption architecture further enhances the IR electric field localization strength, we expect that this integration strategy can  enhance molecular vibration coupling or non-linear optical response. The versatility of the NC assembly and integration approach suggests opportunities for various metal oxide NC superstructures, including mixing and stacking of NCs beyond a single monolayer, representing a vast parameter space for design of linear and nonlinear IR optical components.

\end{abstract}


\section{Introduction}
\begin{figure}
    \centering
    \includegraphics[width=1\textwidth]{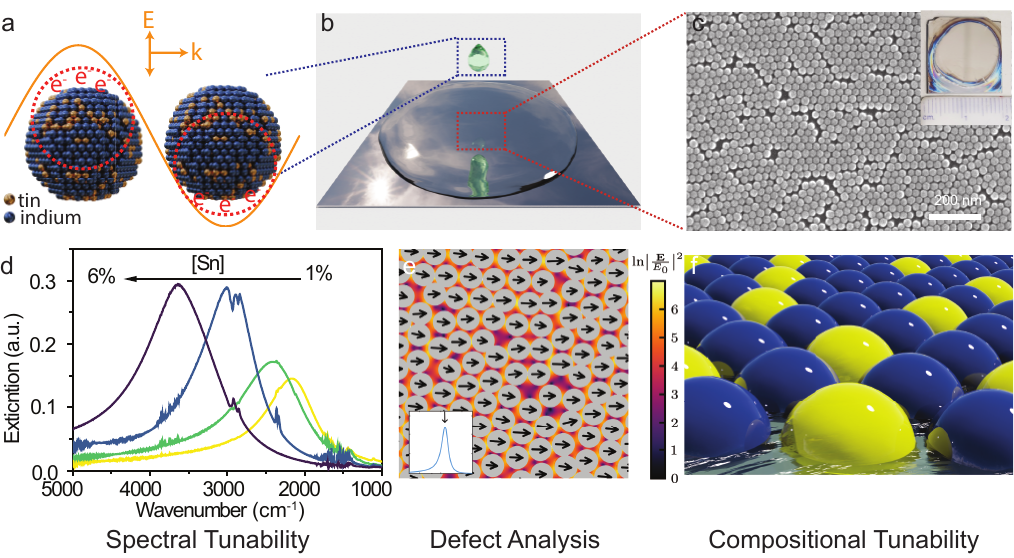}
    \caption{\textbf{Preparation of metasurfaces from tin-doped indium oxide (ITO) nanocrystals.} (a) Schematic of localized surface plasmon resonance in an ITO NC. $E$ and $k$ indicate the electric field and the wavevector of the incident light. (b) Illustration of liquid-air interface assembly method for NC monolayer preparation. NCs in a nonpolar solvent are drop-cast onto a polar antisolvent subphase. (c) Scanning electron microscope image of an ITO NC monolayer. The inset shows the scalability of the process, resulting in a NC monolayer covering a few square centimeters. (d) Extinction spectra of monolayers of ITO NCs (around 30 nm in diameter) on silicon with Sn doping concentrations from 1 to 6\%. (e) Simulated near-field optical response of a defective NC monolayer. Arrows indicate the direction of the imaginary dipole. The inset image indicates the probe frequency. (f) Schematic of a NC monolayer assembly incorporating two different Sn doping concentrations. Panels b, c, f adapted with permission from Reference~\cite{kim2023hierarchically}; copyright 2023 American Chemical Society. Panel e adapted with permission from Reference~\cite{green2024structural}; copyright 2024 American Chemical Society.}\label{Fig:fig1}
\end{figure}

Optical metamaterials can manipulate the pathways and material interactions of electromagnetic waves through their engineered permittivity and permeability responses that extend beyond those found in naturally occurring bulk materials.\cite{valentine2008three,cai2010optical} These metamaterials typically consist of a periodic arrangement of sub-wavelength nanosized unit cells and can incorporate metals and dielectric materials as building blocks. There is growing interest in tuning these metamaterials at infrared (IR) frequencies due to the presence of numerous molecular absorption resonances and optical phonons of various polar materials in the IR spectrum, as well as the importance of IR light in a wide range of applications.\cite{jiang2011conformal} IR metamaterials could be used for thermal control,\cite{li2024use} telecommunications,\cite{dolling2006low} stealth coatings,\cite{kim2017selective} molecular sensing,\cite{wang2022molecular} and more. A particular focus is the development of IR resonant metasurfaces that strongly interact with light despite being very thin, which can increase their sensitivity to environmental signals for sensing\cite{SLawNL} and enable the strong nonlinear effects required for harmonic generation and optical modulation.\cite{li2017nonlinear}

To achieve these goals, top-down approaches, such as electron-beam or photolithography, have been employed to create metamaterial structures by etching or additive patterning of metallic components. Various geometries of periodic unit cells, including nanorods,\cite{aksu2010high} nanorings,\cite{halpern2013lithographically} and more complex geometries have been created, resulting in a rich variety of optical phenomena. However, the energy and cost requirements for top-down patterning are substantial, and complicated processing and slow patterning compromise the potential for cost-effective technologies, especially for applications targeting large surface areas, such as light-driven energy storage\cite{ashrafi2024elliptical} or optical coatings for thermal control.\cite{Mason:10,guo2022real} Additionally, the bottleneck of expensive lithographic processing worsens when smaller or more intricately structured unit cells are required, which can be necessary to maximize the concentration of the electric near-field within these metamaterials.

As an alternative, the periodic patterns for metamaterials can be generated by arranging colloidal metallic nanocrystals (NCs) through self-assembly, where each NC serves as a basic unit cell.\cite{chen2019designing} With the proper assembly method, a periodic structure can be prepared in a more scalable manner, surpassing areas of a few centimeters even with simple lab-scale processing and potentially scaling to meters-wide substrates with high throughput coating processes. Moreover, the colloidal synthesis of NCs and their assembly into periodic structures do not require high vacuum or demanding instrumentation, further contributing to the scalability of this approach. For an assembled superstructure of metallic NCs, the optical response results from the collectively coupled plasmon resonance (CPR) of the localized surface plasmon resonances (LSPRs) of individual NCs. Therefore, the tunability of individual NC LSPR frequencies can provide flexibility in CPR frequency as well as the permittivity function of the NC superstructure, which enables tuning of the optical response for desired applications.

However, several fundamental challenges arise when using conventional metallic NCs for IR-resonant metamaterials. First, since each metal has a fixed charge carrier concentration ($\sim10^{22}~\text{cm}^{-3}$), the primary way to tune LSPR frequency is by changing the shape and size of NCs (\textit{e.g.}, nanorods)\cite{li2018infrared}, requiring nanostructures with long axis of several hundred nanometers or more for IR resonance.\cite{law2013towards,maier2006terahertz} Such large NCs have poor colloidal stability, hindering the ability to make uniform and scalable NC assemblies by solution processing.\cite{li2018infrared} Alternatively, integrating smaller NCs into optical cavity architectures or precisely stacking them into multiple layers can shift the visible range LSPR to IR resonant CPR by coupling to photonic modes.\cite{chang2023wavelength,arul2022giant,mueller2020deep} Besides limiting the scalability of fabrication methods, using large NCs or stacking multiple layers of NCs also diminishes the spatial concentration of the localized optical modes (\textit{i.e.}, increases the mode volume). Considering that many compelling applications of metamaterials stem from their strong field localization, a larger mode volume is detrimental in designing metamaterials.

Doped metal oxide NCs are suitable candidates for superlattice building blocks that circumvent some of the drawbacks of conventional metallic NCs. These materials decouple size from frequency response since they exhibit an IR LSPR frequency even for small NC diameters (Figure~\ref{Fig:fig1}a) due to their low charge carrier concentrations ($\sim10^{19}-10^{21}~\text{cm}^{-3}$).\cite{agrawal2018localized} The electron concentration that governs the LSPR frequency is tunable since these free electrons are introduced by doping the host materials, which are wide-bandgap semiconductors. Unlike conventional metals, the large bandgaps in the UV region make them selectively responsive to IR frequencies, enabling visibly transparent optical components and avoiding any parasitic interaction with visible light. These features of metal oxide NCs provide sweeping flexibility in designing and fabricating NC superstructures as IR resonant metamaterials. 

While doped metal oxide thin films have been studied as tunable linear and nonlinear optical materials,\cite{alam2016large,he2021deterministic} relatively few studies have been published on assembled metal oxide NC structures. The additional tunability offered by metal oxide NC films and the presence of electromagnetic hot spots in the gaps between NCs suggests that the bottom-up approach holds promise for novel metamaterial applications, especially at IR frequencies. Therefore, our research groups have been focused on understanding the optical response of metal oxide NC monolayers and the potential applications of these metamaterials in energy, optics, and information technology. Among various compositions of metal oxides, our Account focuses on tin-doped indium oxide (ITO) NCs, where the electron concentration and bulk plasma frequency is controlled by the tin concentration.\cite{gibbs2020intrinsic} Highly uniform NC shapes with precisely tuned size and doping concentration are achieved by the slow injection method introduced by the Hutchison group.\cite{jansons2016continuous} Guided by the liquid-air interface (Figure~\ref{Fig:fig1}b), a compact NC monolayer can be formed, which is transferable to arbitrary substrates.\cite{dong2010binary} This method allows uniform and scalable assembly of the NCs, regardless of their size and composition (Figure~\ref{Fig:fig1}c). In such well-controlled NC monolayers, we demonstrated a resonant optical response that concentrates IR light strongly in the nanoscale gaps between the NCs at frequencies that can be experimentally tuned based on the size and doping concentration in the NCs (Figure~\ref{Fig:fig1}d, e). Independent control over the amplitude of the extinction and the resonance frequency by synthetically adjusting the NC size and Sn doping concentration facilitates metamaterial properties optimization.\cite{staller2019quantitative}

The impact of structural defects, an apparent limitation of the bottom-up approach, was revealed by large-scale simulations (Figure~\ref{Fig:fig1}e) to be moderate and even advantageous for applications like surface-enhanced Raman spectroscopy (SERS) that rely on the intensity of near-field enhancement (NFE).\cite{sherman2023plasmonic, green2024structural} Meanwhile, the flexibility of self-assembly to incorporate disparate compositions of NCs (Figure~\ref{Fig:fig1}f) and molecules within the metasurface is a great advantage over top-down fabrication.\cite{kim2023hierarchically,chang2024surface,sherman2023plasmonic} Finally, the facile integration of self-assembled NC monolayers with photonic structures is demonstrated with a simple cavity architecture that achieves 100\% IR light absorption across a dopant-tunable wavelength range.\cite{chang2023wavelength} From these foundational insights to potential applications, we establish doped metal oxide NC monolayers as metamaterials that have the potential to contribute to sustainability by managing heat, advance information science through non-linear optical responses, and enhance environmental science by detecting molecular entities.

\subsection{Impact of Structural Defects in NC Monolayers}

\begin{figure}
    \centering
    \includegraphics[width=1\textwidth]{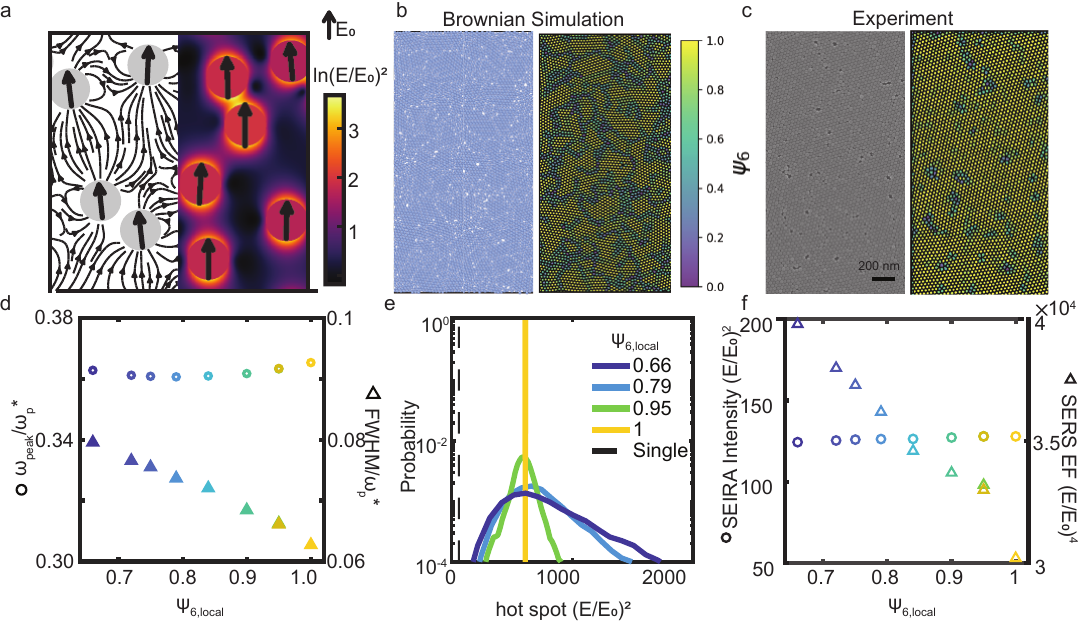}
    \caption{\textbf{Impact of disorder on optical properties.} (a) Electric field map for a configuration of ITO NCs generated using MPM. (b) Defective NC monolayer simulated with Brownian dynamics (left, particles shown in blue) and the corresponding local bond-orientational order $\psi_6$ (right). (c) SEM image of an ITO NC monolayer (left) and the corresponding local $\psi_6$ values captured by a deep-learning-based algorithm. (d) Normalized resonance peak frequency ($\omega_\text{peak}$) and full-width-at-half-maximum (FWHM) from the far-field spectra calculated using MPM vs average $\psi_6$. (e) Hot spot intensity distributions vs average $\psi_6$. (f)  Predicted SEIRA and SERS EF dependence on average $\psi_6$. Panel a adapted with permission from Reference \cite{sherman2023plasmonic}; copyright 2023 American Chemical Society. Panels b-f adapted with permission from Reference \cite{green2024structural}; copyright 2024 American Chemical Society.}\label{Fig:fig2}
\end{figure}

Self-assembly inevitably results in defects, which are readily observed by scanning electron microscopy (SEM) of NC monolayers (Figure~\ref{Fig:fig1}c). Understanding and predicting the impact of these defects on optical properties is essential, especially for developing applications at scale. Conventionally, finite element method (FEM) or finite-difference time-domain methods are used to compute the optical properties of model NC assemblies, providing accurate predictions for small systems of nanoparticles of arbitrary shape. These methods are well-suited to simulate perfectly ordered structures, including defect-free NC monolayers. However, the unit cell required to represent realistic disorder in a NC monolayer is large, and such methods become computationally intractable.

To address these challenges, our groups developed a mutual polarization method (MPM), which rapidly solves Maxwell's equations for the optical response in large systems of arbitrarily arranged spherical particles in the quasistatic limit (Figure~\ref{Fig:fig2}a).\cite{sherman2023plasmonic,sherman2024distribution} This method builds on earlier coupled dipole methods\cite{ross2015defect} and efficiently calculates the dipole and quadrupolar moments within each particle, as influenced by the other $N-1$ particles in the system via a linear system of equations. MPM can calculate the response of complex systems with periodically replicated unit cells of more than $10^4$ particles. Since MPM is tolerant to minor particle overlaps that can occur in simulated self-assembly processes, it is particularly suited for assessing the optical implications of structural defects in dense assemblies like grain boundaries, vacancies, or other disorder. 

To investigate the impact of defects, we first simulated NC monolayers with varying degrees of disorder by compressing a Brownian dynamics simulation cell of 8,100 NCs at varying rates, mimicking varying evaporation rates in experimental assembly formation—where slower rates lead to more well-ordered configurations (Figure~\ref{Fig:fig2}b).\cite{green2024structural} We assessed the perfection of NC organization using the hexatic order parameter ($\psi_6$), which describes the relative orientation of virtual ``bonds'' between a NC and its neighbors, compared to a perfect hexagonal lattice ($\psi_6=1$). The particle-averaged $\psi_6$ values for each configuration ranged from 0.6 to 1.0, which spans the range from liquid-like to perfectly ordered monolayers and encompasses the defectivity of experimental NC monolayers made by self-assembly ($\psi_6 \approx0.85$), including ITO NC monolayers assembled in our lab and analyzed by a deep-learning-based algorithm to find $\psi_6$  (Figure~\ref{Fig:fig2}c). 

By varying $\psi_6$ in Brownian dynamics simulations and predicting the far-field spectra with MPM, we observed that the peak position of CPR is relatively insensitive to $\psi_6$, although the linewidth of the spectrum broadens somewhat as $\psi_6$ decreases (Figure~\ref{Fig:fig2}d). The applicability of these observations to experimental ITO NC monolayers was validated by comparing their measured spectra with the MPM-calculated extinction spectra of monolayers having similar $\psi_6$. When using the experimentally deduced NC dielectric functions\cite{gibbs2020intrinsic} as an input to MPM, the spectra are well reproduced, supporting the use of MPM to further investigate and predict optical properties of NC assemblies, including those with complex structures and defects.



Although far-field spectra only show modest changes in linewidth with structural defectivity, disorder has a pronounced effect on the distribution of near-field enhancement (NFE). We observed a wider distribution of particle hot spot NFE factors with lower $\psi_6$ (Figure~\ref{Fig:fig2}e). As increasing disorder is introduced, some NCs experience more strongly enhanced local electric fields, an effect which correlates with a red shift of their local resonance frequency. These stronger electric field hot spots can often be found in a ``chain'' of NCs aligned parallel to the field polarization of the incident light and separated by relatively smaller gaps compared to perfectly ordered NCs.\cite{sherman2024illuminating}

Since the NFE is responsible for enhancing the sensitivity of spectroscopic detection of molecules, we evaluated the impact of disorder on ensemble-averaged enhancement factors (EFs) for surface-enhanced infrared absorption (SEIRA) and SERS (Figure~\ref{Fig:fig2}f). For these calculations, we averaged the electric field enhancement on the surface of NCs, where the probed molecules are presumed to be distributed randomly. For the SEIRA EF, which depends on the intensity of the electric field around the molecules relative to the incident electric field intensity, expressed as $|E/E_0|^2$, a relatively consistent value was achieved, regardless of average $\psi_6$ in the assembly. This consistency reflects the averaging of contributions from hot spots with higher and lower EF and indicates significant defect tolerance for SEIRA applications. The SERS EF, which is expected to scale as $|E/E_0|^4$, grows with increasing disorder (decreasing $\psi_6$), consistent with prior results that also predicted that defective NC superlattices will have increased EF and give rise to stronger SERS signals.\cite{ross2015defect} This behavior derives from the greater contribution to SERS by the subpopulation of hot spots with a larger than average NFE. This analysis shows that self-assembled metamaterials with colloidal NC building blocks can be designed and fabricated where inevitable defects do not diminish---and potentially even enhance---the optical sensing properties.

While our computational assessment of disorder focused on ITO NCs, the approach is material-agnostic, and alternative compositions or more complex geometries can be considered. The permittivity of each particle or sub-population of particles is provided as an input and arbitrary structural configurations can be created to represent different candidate fabrication strategies. For particles with a Drude-like dielectric response, we demonstrated that sensitivity of optical properties to structural disorder is greater for increased dipole polarizability.\cite{green2024structural,sherman2024distribution} 
We foresee that by expanding beyond the quasistatic limit, we could apply MPM to three-dimensional superlattices, including additional defect types and compositional mixtures. Furthermore, owing to its computational efficiency, MPM should be useful for design, including solving inverse problems where particle positions or dielectric functions in an assembly are optimized using deep learning-based strategies to achieve a target far-field or near-field optical response of the metamaterial.\cite{kadulkar2022machine} 

\subsection{Vibration Signal Enhancement in ITO NC Monolayers}
\begin{figure}
    \centering
    \includegraphics[width=1\textwidth]{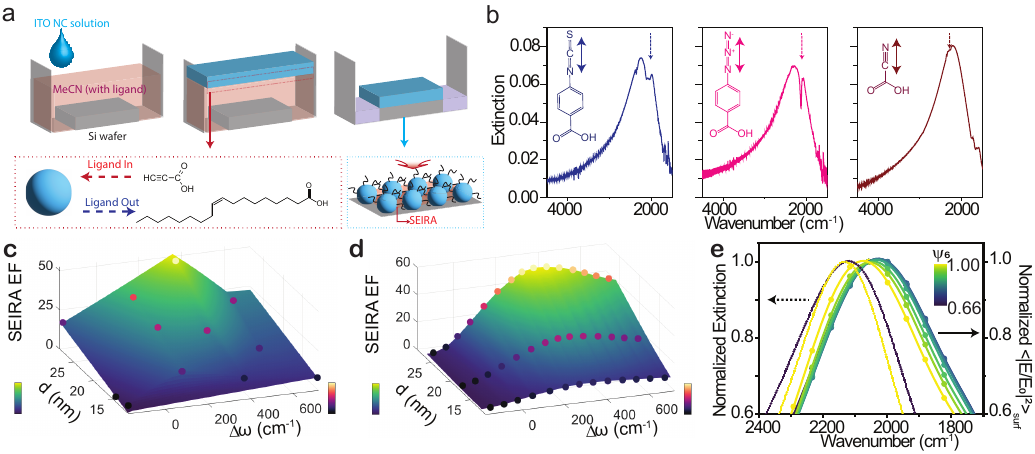}
    \caption{\textbf{Spectroscopic detection of molecular vibrations coupled to the CPR of ITO NC monolayers.} (a) Schematic of NC monolayer preparation with ligand exchange to incorporate vibrational probe molecules. (b) Extinction spectra for ITO NC monolayers with various molecular vibrations appearing as induced transparency features. (c) Surface-enhanced infrared absorption intensity enhancement factor (SEIRA EF) as a function of NC diameter and detuning, $\Delta\omega$, controlled by the Sn doping concentration. (d) SEIRA EF trends calculated with the finite element method. (e) Extinction spectra and near-field enhancement intensity on the surface of NCs ($\langle|E/E_0|^2\rangle_\text{surf}$) as a function of $\psi_6$, calculated using MPM. Panels a-e adapted with permission from Reference \cite{chang2024surface}; copyright 2024 American Chemical Society.}\label{Fig:fig3}
\end{figure}

The strongly localized IR electric field generated in the gaps within ITO NC monolayers can couple the light efficiently to incorporated molecules, specifically to their resonant vibrational modes.\cite{chang2024surface,agrawal2017resonant,Xi2018} This coupling is the basis for SEIRA, which enhances the signal intensity for molecular vibrations and can enable chemically specific detection at low concentrations. Since each molecular vibration response occurs at at a specific frequency, the NC dopant concentration, which controls the peak frequency of the CPR in the NC monolayer, should be optimized. Additionally, the size of the NCs influences the absorption-to-scattering ratio and the strength of coupling between the LSPRs, rendering it another tunable parameter to be considered. This parameter space allows optimization of the SEIRA effect by maximizing the electric field seen by the molecule at frequencies overlapping with the molecule's vibrational resonances and enables a mechanistic evaluation of the conditions for optimal SEIRA. 

While preparing a NC monolayer, we can incorporate molecules containing the vibrational modes of interest through a ligand exchange process  (Figure~\ref{Fig:fig3}a).\cite{chang2024surface,dong2013electronically} The probe molecules are dissolved in the polar subphase and they partially replace the native oleate ligands on the surface of the NC, causing contraction of the interparticle spaces. Thus, the probe molecules remain embedded in the gaps when the monolayer is transferred to a substrate for analysis.\cite{dong2013electronically} With this method, we incorporated molecules bearing various chemical moieties into the assemblies while maintaining a consistently compact NC monolayer, regardless of size and Sn doping concentration in the NCs. When the extinction spectra were measured, in addition to the broad plasmon feature, we observed a sharp dip at the molecular vibration frequency (Figure~\ref{Fig:fig3}b). Considering that molecular vibrations are absorptive and would typically appear as a peak, this dip, which is called an electromagnetically induced transparency, is a clear indication of coupling between the CPR and molecular vibrations.\cite{adato2013engineered} This coupling is not limited to a particular vibrational mode but rather occurs generally for molecules that incorporate binding groups enabling adsorption onto the NC surface, as carboxylates do on ITO.

We systematically studied the experimental SEIRA EF, which is defined as the vibrational intensity relative to the intensity observed for non-plasmonic undoped \ch{In2O3} NCs (Figure~\ref{Fig:fig3}c). The signal intensity was quantified by first isolating the vibrational signal from the broad plasmon background, after which the Fano-like lineshape was apparent.\cite{chang2024surface} The maximum SEIRA EF observed was as high as $50\times$, which is the highest value reported for metal oxide-based SEIRA platforms and also comparable with metallic NC-based SEIRA platforms for molecular vibration enhancement. In the size range studied, SEIRA EF increased with the NC diameter. On the other hand, a non-monotonic trend was noted as a function of detuning ($\Delta\omega$), which is the difference between the molecular vibration frequency ($\omega_\text{mol}$) and the peak frequency of the NC monolayer CPR ($\omega_\text{CPR}$) that we controlled by synthetically adjusting the Sn doping concentration. Interestingly, the maximum SEIRA EF was achieved at positive detuning rather than perfectly on resonance.

We employed simulations using FEM and MPM to elucidate the mechanisms behind the detuning and NC size-dependent SEIRA EF trends (Figure~\ref{Fig:fig3}d). We learned that vibrational intensity is influenced by the combination of plasmon polarization strength (which is greater for larger and more highly doped NCs) and the frequency dependence of the NFE. We found that the NFE response is spectrally offset from the far-field spectrum, with its maximum intensity being red-shifted in frequency even for perfectly ordered monolayers. From MPM simulations, we discovered that the redshift is magnified with increasing disorder (Figure~\ref{Fig:fig3}e). The experimental realization and computational analysis of SEIRA in ITO NC monolayers confirms and elaborates on our prediction that these metamaterials have useful NFE characteristics that are tunable and defect-tolerant. The size and doping concentration of the NCs are useful handles to optimize and customize the metamaterials for applications like molecular sensing. We envision these NC monolayers could be integrated into flow-through sensors for detecting the presence of target molecules in waste streams or as biomarkers.

The design principles gleaned from studying SEIRA trends are informative beyond vibrational sensing applications. We showed that the molecular vibration intensity is a useful probe of NFE within NC-based metamaterials. Looking forward, we expect that vibrational probe molecules can be used in additional assembly configurations of NCs and in plasmonic-photonic hybrid structures for experimentally reporting on the extent of local field enhancement. In addition, the strategy for coupling NC-based CPRs and molecular vibrations could be extended to other material components. For example, metasurface-embedded molecules or quantum dots could benefit from the high local NFE to enable efficient two-photon fluorescence. The transparency of the ITO NCs to higher frequency light would allow efficient out-coupling that would be inhibited by metamaterials based on conventional metal NCs.

\subsection{Epsilon Near Zero Properties of Statistically Mixed Monolayer Superlattices of ITO NCs}
\begin{figure}
    \centering
    \includegraphics[width=1\textwidth]{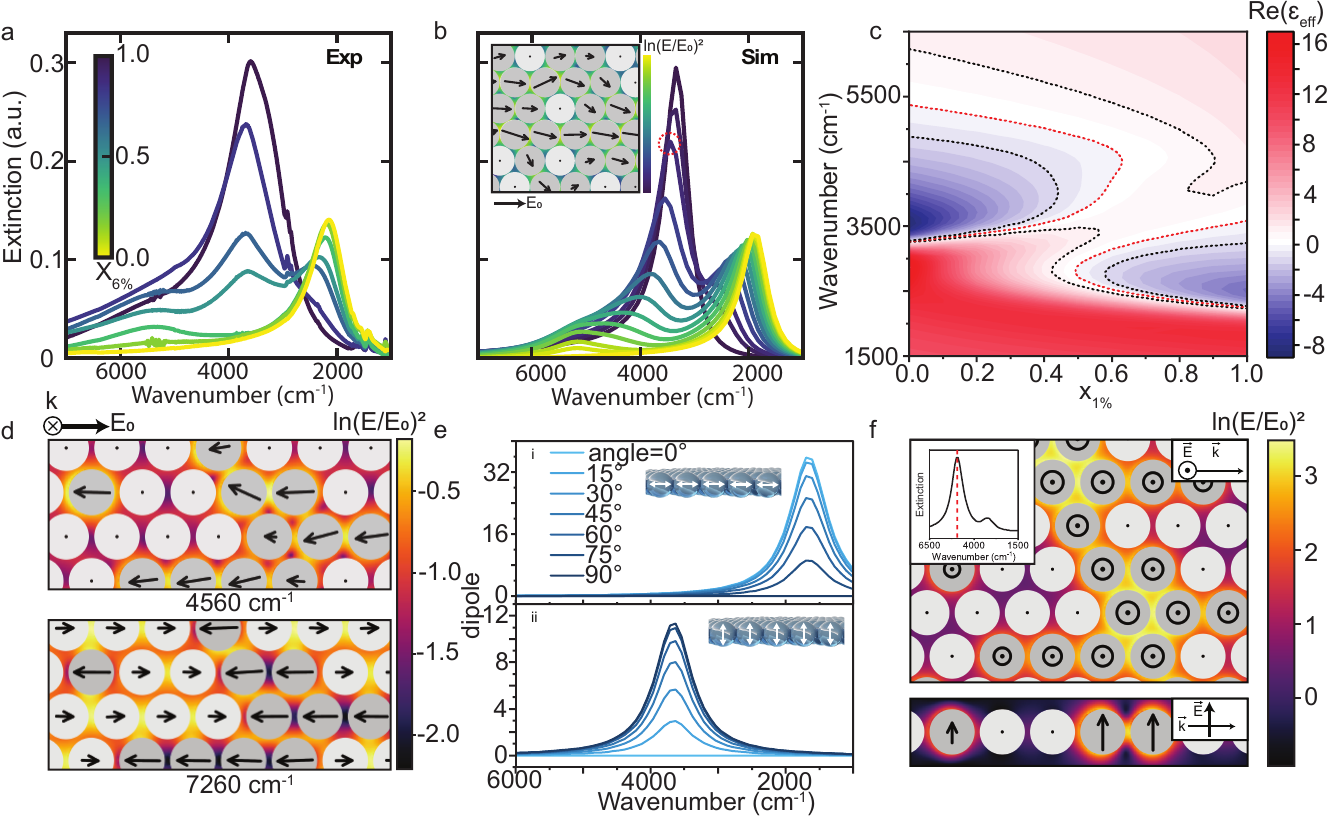}
    \caption{\textbf{Effective permittivity engineering of NC monolayer superlattices composed of a statistical mixture of ITO NCs.} 
(a) Extinction spectra of NC monolayers with 30 nm diameter NCs including a statistical mixture of 1\% and 6\% Sn ITO NCs. The percentage here refers to the tin doping concentration, while the mixing fraction is shown as a function of the 6\% ITO NC ratio ($x_{6\%}$). (b) Extinction spectra from MPM simulations with the same compositions as in (a). The inset shows the imaginary part of the dipoles in the mixed ITO NC monolayer with ($x_{6\%}=0.8$) ITO NCs (dark gray). The corresponding frequency is indicated by the red dashed circle on the spectra. (c) Effective real permittivity of ITO NC monolayers composed of a statistical mixture of 1\% and 3\% Sn ITO NCs with 30 nm diameters, calculated by MPM. Permittivity is shown as a function of the 1\% ITO NC ratio ($x_{1\%}$). (d) Real part of the dipoles and local electric field intensity of a $x_{1\%}=0.57$ and $x_{3\%}=0.43$ ITO NC mixed monolayer. (e) Incident angle-dependent imaginary component of the induced dipole intensity of a 1\% Sn ITO NC monolayer with (i) in-plane polarization and (ii) out-of-plane polarization of the electric field. (f) Electric field map of a statistical mixture of ITO NCs ($x_{1\%}=0.57$ and $x_{3\%}=0.43$) with the out-of-plane polarization shown top-down and in cross-section (below). Panels a,b adapted with permission from Reference~\cite{sherman2023plasmonic}. Panels c-f adapted with permission from Reference~\cite{kim2023hierarchically}; Both copyrights 2023 American Chemical Society.}\label{Fig:fig4}
\end{figure}

In addition to enabling straightforward incorporation of molecules of interest, the self-assembly of NC-based metasurfaces readily facilitates the mixing of compositionally distinct NC components. An arbitrary statistical mixture of ITO NCs with two different dopant concentrations can be prepared in solution, which directly determines the composition of the corresponding mixed NC monolayer superlattice (Figure~\ref{Fig:fig1}e).\cite{sherman2023plasmonic,kim2023hierarchically} The facile intermixing produces compositionally complex metamaterials with nanoscale delineation of the components that would be nearly impossible to realize by lithographic means. The NCs remain intermixed and do not phase separate during assembly because the surface chemistry is very similar for indium oxides, regardless of the doping concentration of the NCs (the Sn concentration is usually between 0-10\%), the Hamaker constant is effectively dopant independent\cite{bodnarchuk2010energetic}, and the size of the two compositions of ITO NCs is the same. Although the preparation of these mixed-composition metasurfaces is trivially different than their single-component counterparts, their optical properties are markedly affected by non-trivial cross-coupling between the components. For example, in statistically mixed monolayers of ITO NCs with 1\% and 6\% Sn concentrations, we observed component peak intensities and positions in the far-field extinction spectra that deviate strongly from linear mixing as a function of the NC ratio (Figure~\ref{Fig:fig4}a).\cite{sherman2023plasmonic} For example, the peak height at low frequency is stronger than expected and the peak at high frequency is weaker than expected based on linear mixing. Using MPM, which accounts for inter-component coupling, we accurately reproduced the experimental trends, using configurations of 11,600 NCs to faithfully represent the statistically random mixing. The near-field maps generated using MPM help explain the non-proportional mixing behavior: high and low-doped NCs mutually contribute to polarization at the low-frequency peak, while the low-doped NCs do not contribute to boosting polarization at the high-frequency peak (Figure~\ref{Fig:fig4}b, inset). Therefore, this difference in coupling from high- and low-doped NCs, evident in the near-field optical behavior, must be considered in predicting or rationalizing the far-field response.

Because the NCs are much smaller than the wavelength of IR light, statistical mixtures can also be used to tune the effective permittivity, $\varepsilon_{\text{eff}}$, of their monolayer assemblies. A monolayer superlattice containing a single composition of ITO NCs exhibits a Lorentzian-like permittivity response where the imaginary part [Im($\varepsilon_\text{eff}$)] follows the Lorentzian lineshape and the real part [Re($\varepsilon_\text{eff}$)] resembles the derivative of a Lorentzian. Just like the extinction spectra, the dielectric response of mixed NC superlattices deviates from a linear combination of these Lorentzian lineshapes due to cross-coupling between dissimilar composition NCs.\cite{kim2023hierarchically}

To demonstrate the possibilities for designing metasurfaces from mixed NC monolayers, we prepared monolayers containing mixed 3\% ITO and 1\% Sn ITO NCs. Over a range of frequencies, the Re($\varepsilon_\text{eff}$) of a homogeneous 1\% Sn ITO NC monolayers is opposite in sign to the Re($\varepsilon_\text{eff}$) of a homogeneous 3\% Sn ITO NC monolayer. Therefore, in the statistical mixtures, we hypothesized that the two components might have an offsetting effect on the Re($\varepsilon_\text{eff}$), resulting in a region with near-zero permittivity. We consider Re($\varepsilon_\text{eff}$) between -1 to 1 to be defined as the epsilon-near-zero (ENZ) region, where the polarization response of the medium becomes smaller than that of air, which can result in strong field concentration and exotic optical responses.\cite{AdamsENZ, reshef2019nonlinear} We observed in MPM-based simulations that Re($\varepsilon_\text{eff}$) becomes flat near the frequency region where the two components have an offsetting response (Figure~\ref{Fig:fig4}c). Specifically, with a mixture of 0.57 1\% Sn ITO and 0.43 3\% Sn ITO NCs, the in-plane ENZ frequency region expanded to cover 2500-5500 $\text{cm}^{-1}$, which is more than three times wider than the ENZ region obtained from a pure 3\% or 1\% Sn ITO NC film with the same structure. 

The mechanism underlying the effective ENZ response can be interpreted using the near-field maps from MPM, which show that in the ENZ frequency region, the polarization directions of 3\% and 1\% Sn ITO NCs are opposing each other, leading to a net polarization near zero (Figure~\ref{Fig:fig4}d). Within the ENZ region, the polarization strength of each component varies with frequency but they remain largely offsetting. This antiparallel polarization mechanism differs dramatically from the abrupt reversal of polarization direction for every NC at a single ENZ frequency in single-component NC assemblies, which is similar to the polarization behavior of thin films as they pass through the ENZ frequency.\cite{reshef2019nonlinear,alam2016large} Thus, NC intermixing broadens the ENZ frequency range, potentially enabling broadband nonlinear optical enhancement, as well as broadband directional thermal emission.\cite{xu2021broadband}

At oblique angles of incidence, two distinct optical resonances are observed, even for ITO NC monolayers containing a single composition of NCs. Based on simulations and consistent with the variation of their intensities as a function of incidence angle, we assign the lower and higher frequency features as in-plane and out-of-plane polarized resonances, respectively (Figure~\ref{Fig:fig4}e).  Similar to the in-plane resonances, the out-of-plane resonances of the two different compositions of ITO NCs can also interact with each other in statistically mixed NC monolayers and strong NFE was observed under out-of-plane polarized excitation (Figure~\ref{Fig:fig4}f), tunable as a function of the mixing ratio. 

The observed generation of electric field hot spots for both in- and out-of-plane polarized excitation suggests the potential to leverage these NC monolayer-based ENZ films for high harmonic generation, specifically to extend the spectral range of ENZ region, and thus enhanced nonlinear response. Wen \textit{et al.} demonstrated that their patterned TiN nanostructured metasurfaces showed very efficient second harmonic generation by exciting at the LSPR, leveraging the spectral overlap of the second harmonic and ENZ modes.\cite{wen2018doubly} Such double resonance may be achieved by different mechanisms in mixed NC metasurfaces, suggesting exciting avenues for future studies. For instance, the broad ENZ range could cover both input and output frequencies within a single NC monolayer, or the in-plane resonance could enhance excitation while the out-of-plane resonance enhances the up-converted emission of an embedded molecular chromophore or quantum dot. With the combination of dopant- and size-tunable NC building blocks and mixing ratio tuning of NC monolayers, the parameter space for optical customization is vast.

\subsection{Tunable Perfect Absorption with ITO NC Monolayers}
\begin{figure}
    \centering
    \includegraphics[width=1\textwidth]{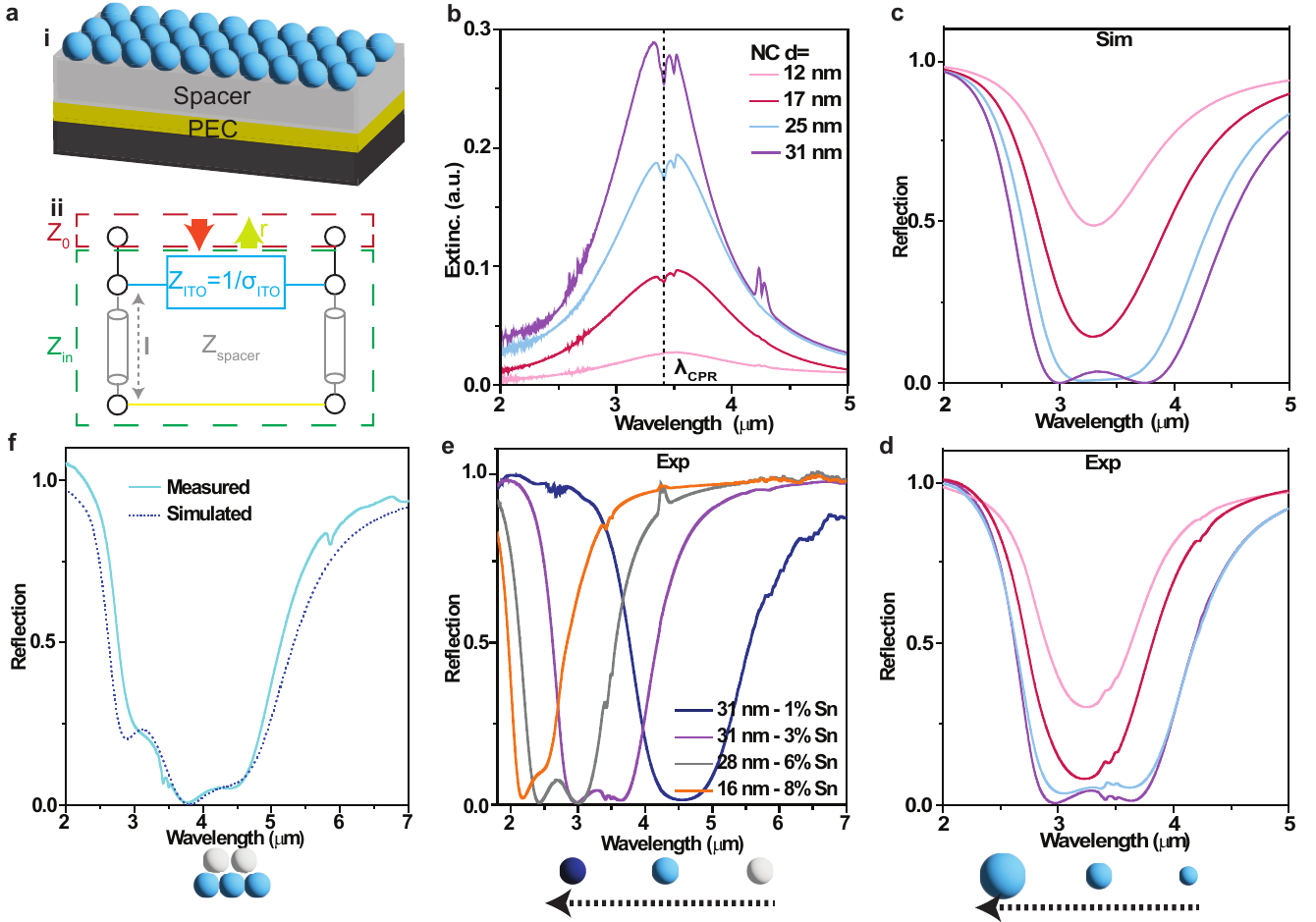}
    \caption{\textbf{Perfect absorption at mid-IR wavelengths by integrating ITO NC monolayers in a photonic structure.} (a) Configuration of NC monolayer on top of spacer and metal (PEC) layers, which is called the Salisbury screen configuration (i). Transmission line theory describing this NC monolayer-based perfect absorber (ii). (b) Extinction spectra of NC monolayers with different sizes of NCs but having the same $\lambda_\text{CPR}$. (c) Simulations from transmission line theory and (d) experimental reflection spectra of different-sized NC monolayers in  the Salisbury screen configuration. (e) Reflection spectra of NCs with different Sn doping concentrations in the Salisbury screen configuration. The spacer layer thickness is optimized for each Sn doping concentration. (f) Experimental and simulated reflection spectrum of a NC bilayer composed of different dopant concentrations in the Salisbury screen configuration. Broad absorption was achieved with this geometry. Panels a-f adapted with permission from Reference \cite{chang2023wavelength}; copyright 2024 American Chemical Society.}\label{Fig:fig5}
\end{figure}

To strengthen interactions beyond the NC metasurfaces themselves, NC monolayers can be integrated into photonic structures. The $\varepsilon_\text{eff}$ of the metasurface can be used for computational design, which we demonstrated by incorporating NC monolayers into a Salisbury screen configuration [Figure~\ref{Fig:fig5}a(i)] to realize IR perfect absorption.\cite{chang2023wavelength} In this architecture, the thickness of a dielectric spacer between the NC layer and an underlying metal reflector is tuned to approximately one-quarter wavelength of the peak wavelength for the Im($\varepsilon_{\text{eff}}$), which is identical to the peak wavelength of the CPR ($\lambda_\text{CPR}$). This geometry positions the antinode of the optical mode in the NC layer to maximize the strength of the light-matter interaction.

This design is based on a recently developed material-agnostic framework described by Sakotic \textit{et al.}, from which we predicted the potential for perfect absorption. In the ultrathin limit, the optical response of the NC layer can be approximated by an effective 2D sheet conductivity $\sigma_{eff}$($\varepsilon_{\text{eff}}$) [Figure~\ref{Fig:fig5}a(ii)], which enables straightforward calculation of reflection and absorption using transmission-line theory. In this regime, the amount of absorption is related to the real part of the sheet conductivity, which is directly proportional to Im($\varepsilon_{\text{eff}}$) and the thickness of the sheet (\textit{i.e.}, NC diameter). At the peak of the Lorentz resonance, which approximates the NC $\varepsilon_{\text{eff}}$ response and where Im($\varepsilon_{\text{eff}}$) has a maximum, there is a critical thickness for which the system absorbs the maximum amount of light. Integrating ITO NC monolayers into a photonic stack with a gold back reflector and germanium dielectric spacer (\textit{i.e.}, Salisbury screen configuration) allowed us to evaluate the prediction of 100\% light absorption, termed perfect absorption, for a metamaterial absorber layer. Achieving 100\% absorption represents the critical coupling condition in one-port (mirror-backed) systems, where the cavity's leakage and dissipation rates are exactly matched. Using the described calculations, it was determined that the diameters of the NCs are comparable to the critical thicknesses derived from their $\varepsilon_{\text{eff}}$.

From MPM simulations of the NC monolayers, we obtained $\varepsilon_{\text{eff}}$ and used it as an input to predict the conditions for perfect absorption with transmission line theory (Figure~\ref{Fig:fig5}c). To test these predictions experimentally, we leveraged the synthetic tunability of ITO NCs to fabricate NC monolayers with varying diameters at a fixed $\lambda_{\text{CPR}}$ (Figure~\ref{Fig:fig5}b). This feat would be unachievable with metallic NCs since the peak would shift unavoidably with size. With ITO NCs, we can compensate for this LSPR coupling-related shift by adjusting the Sn doping concentration so that NCs with distinct LSPR wavelengths in the solution phase assemble into metamaterials with a consistent CPR wavelength. At a fixed $\lambda_{\text{CPR}}$, increasing NC size adjusts the effective absorber thickness, approaching and then exceeding the predicted critical thickness for perfect absorption. Indeed, with smaller ITO NCs, the reflection minima (absorption maxima) were far from zero (one) due to the NC diameter being less than the critical thickness (Figure~\ref{Fig:fig5}c), resulting in what is called undercoupling. Conversely, with 25~nm ITO NCs, due to the critical thickness matching the NC size, perfect absorption was observed. When the diameter exceeds the critical thickness, perfect absorption occurs at shorter and longer wavelengths, away from $\lambda_{\text{CPR}}$, in what is known as overcoupling behavior. This experimental result (Figure~\ref{Fig:fig5}d) agrees with the predicted spectra (Figure~\ref{Fig:fig5}c) based on the framework from Sakotic \textit{et al.}\cite{sakotic2023perfect}

We also varied the Sn doping concentration at a fixed NC size to tune the peak wavelength of Im($\varepsilon_{\text{eff}}$) and achieve perfect absorption at adjustable wavelengths, with the dielectric spacer thickness tuned to match each $\lambda_{\text{CPR}}$ based on the theory\cite{sakotic2023perfect} (Figure~\ref{Fig:fig5}e). From the understanding we established, we further designed a broadband perfect absorber by stacking NC monolayers with different Sn concentrations. By stacking 1\% and 3\% Sn ITO NC monolayers, we achieved more than 80\% light absorption between the wavelengths of 3 to 5 $\mu$m (Figure~\ref{Fig:fig5}f). Our photonic cavity-integrated NC metamaterials maintain strong absorption even at a higher incidence angle, irrespective of polarization, which can facilitate practical applications.\cite{chang2023wavelength}

Although this work primarily addressed the far-field optical behavior, the resulting structures are predicted to yield very high NFE, suggesting additional potential applications. From FEM simulations of the near-field maps, we observed that an ITO NC monolayer incorporated in a Salisbury screen configuration produces a strong electric field ($|E/E_0|^2$) in the $\approx3$ nm NC gaps reaching a magnitude of around $10^{3.5}$ that well exceeds the NFE in a NC monolayer on its own. Based on the observed intense field enhancement, this platform holds promise for nonlinear optical applications. Because of the strong light-matter interaction, the threshold for light intensity that saturates the absorption response may be modest. Once the intensity exceeds that threshold, light would be partially transmitted through the NC monolayer, then reflected by the back reflector, effectively operating as an intensity-threshold switch. Since the plasmon decay is typically fast, this perfect absorber design might be useful for a mode-locked laser with a short pulse duration. Note that mode-locking across the mid-IR frequency range is challenging\cite{ma2019review} due to the requirement for an effective medium and photonic structure with high controllability throughout this range.

The Salisbury screen architecture can achieve various coupling regimes depending on NC size, each suggesting possible avenues for further development. Critical coupling, where 100\% of the light is absorbed is beneficial when full absorption is a virtue, such as for radiative cooling purposes.\cite{wu2018design} In a deliberately overcoupled system, radiative damping becomes more significant than non-radiative damping, and coupling to a dipolar material introduced within the NC gaps can be enhanced. For example, Paggi \textit{et al.},\cite{paggi2023over} demonstrated maximal vibrational signal intensity for molecular vibrations coupled to a patterned metal layer in the Salisbury screen configuration in the overcoupled regime. Similarly, the emission intensity of other materials can be engineered with the Salisbury screen geometry, as previously shown by Wang \textit{et al.}\cite{wang2019enhanced} using a metal-based Salisbury screen configuration to induce the Purcell effect in a quantum dot film. The tunability of each component of our NC metamaterials integrated in the photonic structure, including the NC building blocks and the dielectric spacer, allows for the optimization of the far-field linear and nonlinear response and of NFE at specified optical frequencies to target a variety of potential applications.

\subsection{Outlook}

In this Account, we focused on ITO NCs as building blocks in NC monolayers that function as IR metamaterials. We have shown that these metamaterials' far- and near-field properties are defect-tolerant or even enhanced by disorder.\cite{green2024structural,ross2015defect,agreda2023tailoring} Their chemical tunability allows embedding additional optical components, like placing molecules in the gaps where their vibrational modes can couple to the CPR of the NC monolayers. The effective permittivity of the metamaterials is adjustable not only by synthetic control of the NCs but also by the expansive parameter space when considering mixed composition NC assemblies. As we have demonstrated by coupling NC monolayers to a simple photonic cavity, we can consider the NC monolayers themselves as highly tunable ``designer effective media'' for constructing three-dimensional NC superstructures by stacking different compositions of NC layers and integrating them with other thin film materials.\cite{berry2024incorporating,erdem2020thickness,chang2023wavelength} We envision constructing 3D metamaterials layer-by-layer, customizing each layer by including different NC compositions or embedded molecules. For example, colloidal hyperbolic metamaterials could be envisioned that concentrate and redirect light via an anisotropic dielectric response.\cite{cleri2023tunable} To further broaden the range of properties achievable in NC-based metamaterials, compositions beyond ITO NCs, like In-doped CdO NCs,\cite{gordon2013shape,shubert2024depletion} which exhibit a higher Q factor and stronger dipole strength due to their elevated mobility, can be considered. To reach shorter wavelengths in the NIR, cesium-doped tungsten oxide (\ch{Cs:WO3}) NCs are promising building blocks.\cite{kim2016interplay} Since MPM captures NC coupling interactions across length scales, this modeling strategy could be extended and used for inverse design to predict compositions and 3D NC architectures that optimally produce the optical properties of interest.

The properties of metal oxide NC metamaterials could also be dynamically manipulated under electrical or electrochemical bias to shift the CPR frequency and modulate coupling strength between the NCs, as we have shown previously for disordered NC films.\cite{tandon2019competition} The optical modulation results from reversible capacitive charging of the NCs, where electronic filling of the surface depletion layer enhances the coupling between neighboring NCs.\cite{zandi2018impacts} This effect has not yet been applied to design dynamic NC metamaterials, but it could potentially be harnessed for dynamic radiative coolers or IR reflectors.\cite{li2024use} The intense hot spots within the gaps in NC metamaterials have already hinted at opportunities for chemically specific detectors, but further strengthening the coupling to vibrational modes could lead also to modified reactivity of embedded molecules, representing a new strategy for heterogeneous catalysis.\cite{thomas2019tilting} There are many opportunities to explore using metal oxide NCs as foundational building blocks for designing IR metamaterials.

\begin{acknowledgement}
This work was primarily supported by the National Science Foundation through the Center for Dynamics and Control of Materials: an NSF MRSEC under Cooperative Agreement Nos.~DMR-1720595 and DMR-2308817, with additional support from the Welch Foundation (F-1696 and F-1848). The authors also acknowledge support from the Defense Advanced Research Projects Agency (DARPA) under the Optomechanical Thermal Imaging (OpTIm) program (HR00112320022), the Multidisciplinary University Research Initiative of the Air Force Office of Scientific Research (AFOSR MURI Award No. FA9550-22-1-0307). 
\end{acknowledgement}
\section*{Competing interests}
The authors declare no competing interests.

\section*{Biographies}

{\bf Woo Je Chang} received a B.S. and M.S. in Materials Science and Engineering and Bioengineering from Seoul National University, and a Ph.D. in Materials Science and Engineering from Northwestern University. He is currently a postdoctoral scholar in the Department of Chemical Engineering at the University of Texas at Austin. His research focuses on the assembly of nanocrystals for applications in photonic structures and metasurface design.

{\bf Allison M. Green} received her B.S. in Chemical Engineering from the University of California, Berkeley, and a Ph.D. in Chemical Engineering from the University of Texas at Austin. Her research interest focuses on the structure and optical properties of nanocrystal assemblies.

{\bf Zarko Sakotic} received B.S. and M.S. degrees in Electrical Engineering from the University of Novi Sad and a Ph.D. in Physics from the same institution. He is currently a postdoctoral scholar in the Department of Electrical and Computer Engineering at the University of Texas at Austin. His research is focused on physics and applications of infrared light-matter interaction.

{\bf Daniel Wasserman} received his Sc.B. in Engineering/Physics and History from Brown University and his M.S. and Ph.D. in Electrical Engineering from Princeton University. Following appointments at the University of Massachusetts Lowell and the University of Illinois Urbana-Champaign, he joined the faculty in the Department of Electrical and Computer Engineering at the University of Texas at Austin. His research focuses on infrared optics and photonic structure design.

{\bf Thomas M. Truskett} received his B.S. from the University of Texas at Austin and his M.A. and Ph.D. from Princeton University, both in chemical engineering. After pursuing postdoctoral studies at the University of California, San Francisco, he joined the University of Texas at Austin. His research focuses on using statistical mechanics to study soft matter and hybrid materials.

{\bf Delia J. Milliron} received her A.B. in Chemistry from Princeton University and her Ph.D. from the University of California, Berkeley. Following appointments at IBM Research and Lawrence Berkeley National Laboratory, she joined the faculty at the University of Texas at Austin. Her research explores the optical, electronic, and electrochemical properties of inorganic nanocrystals and their processing into functional materials.

\providecommand{\latin}[1]{#1}
\makeatletter
\providecommand{\doi}
  {\begingroup\let\do\@makeother\dospecials
  \catcode`\{=1 \catcode`\}=2 \doi@aux}
\providecommand{\doi@aux}[1]{\endgroup\texttt{#1}}
\makeatother
\providecommand*\mcitethebibliography{\thebibliography}
\csname @ifundefined\endcsname{endmcitethebibliography}  {\let\endmcitethebibliography\endthebibliography}{}



\end{document}